\documentclass[a4paper]{jpconf}
\usepackage{amsmath}
\usepackage{graphicx}
\begin{document}
\title{Achieving high baryon densities in the fragmentation regions in heavy ion collisions at top RHIC energy}

\author{Ming Li and Joseph I. Kapusta}

\address{School of Physics and Astronomy, University of Minneosota, Minneapolis, MN 55455, USA}

\ead{ml@physics.umn.edu, kapusta@physics.umn.edu}

\begin{abstract}
 Heavy ion collisions at extremely high energy, such as the top
energy at RHIC, exhibit the property of
transparency where there is a clear separation between the almost
net-baryon-free central rapidity region and the net-baryon-rich
fragmentation region. We calculate the net-baryon rapidity loss and
the nuclear excitation energy using the energy-momentum tensor obtained
from the McLerran-Venugopalan model. Nuclear compression during the
collision is further estimated using a simple space-time picture. The
results show that extremely high baryon densities, about twenty times
larger than the normal nuclear density, can be achieved in the
fragmentation regions. 
\end{abstract}

\section{Introduction}
In 1980 Anishetty, Koehler and McLerran \cite{Anishetty1980} studied the fragmentation regions in heavy ion collisions at extremely high energy. They argued for the possible formation of quark-gluon plasma in these regions with an energy density of about 2 GeV/fm$^3$ enhanced by a nuclear compression factor of 3.5.  
However, after Bjorken's work in 1983 \cite{Bjorken1983}, people's interests have been mainly focused on the central rapidity region where even larger energy density could be achieved. This nearly net-baryon-free region is also easy to be probed experimentally. The fragmentation regions are only occasionally investaged since then \cite{Csernai1984,Gyulassy1986, Mishustin2002, Frankfurt2003, Mishustin2007}. We report our work on the high baryon densities in the fragmentation regions. To be specific, we use the glasma energy-momentum tensor obtained in the McLerran-Venugopalan model \cite{McLerran1994} to calculate the baryon rapidity loss and the nuclear excitation energy. We then use the same space-time picture of collision in \cite{Anishetty1980} to estimate the nuclear compression factors and obtain the high baryon densities distributed in the fragmentation regions.
 
\section{Baryon Rapidity Loss and Nuclear Excitation Energy}
We consider central collisions of Au-Au nuclei at top RHIC energy $\sqrt{s_{NN}} = 200\, \rm{GeV}$.  We study the simplified 1+1 dimensional motion with transverse motion ignored.  Approximate the nucleus as a collection of slabs with unit surface area locating at different $\mathbf{r}_{\perp}$.  Taking one pair of the colliding slabs as an example, the four-momentum of the projectile slab is $\mathcal{P}_{P}^{\mu} = (\mathcal{E}_{P}, 0 ,0,\mathcal{P}_{P})$. As the slabs recede, the loss of the energy and momentum from the slabs should be equal to the increase of the energy and momentum of the glasma in the regions swept by the slabs. The differential energy momentum conservation should be satisfied $d\mathcal{P}_{P}^{\mu} = -T^{\mu\nu}_{\rm{glasma}} d\Sigma_{\nu}$ where $d\Sigma_{\nu} =(dz, 0 ,0 ,-dt)$ is the differential hypersurface vector.

\begin{figure}[h]
\begin{minipage}{0.5\textwidth}
\centering
\includegraphics[width=0.9\linewidth, height=0.23\textheight]{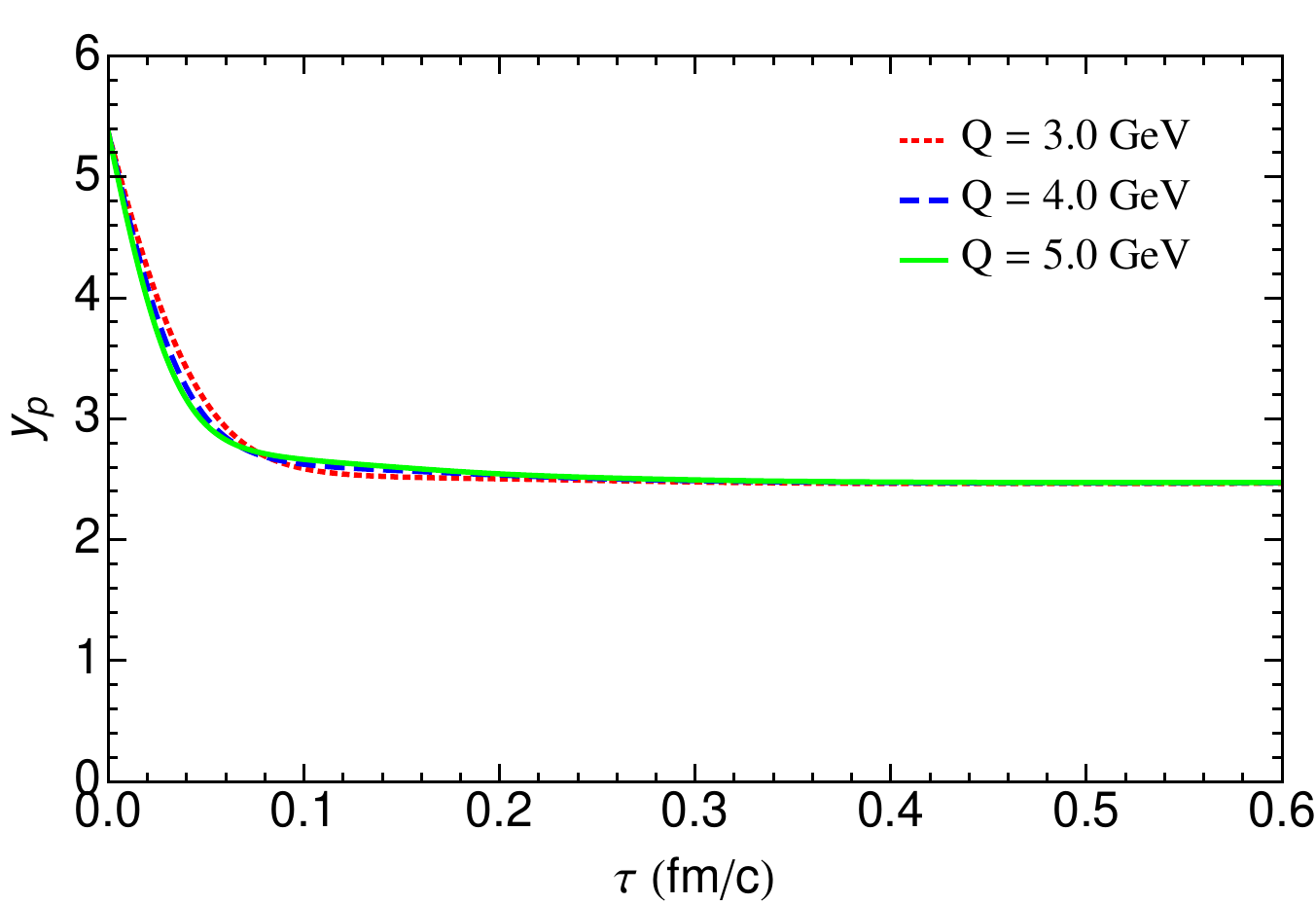}
\caption{\small Projectile rapidity as a function of proper time for the central core slab of the nucleus after the collisions viewed in the center-of-mass frame. }
\label{fig1_y_tau}
\end{minipage}
\hspace{0.1in}
\begin{minipage}{0.5\textwidth}
\centering
\includegraphics[width=0.9\linewidth, height=0.23\textheight]{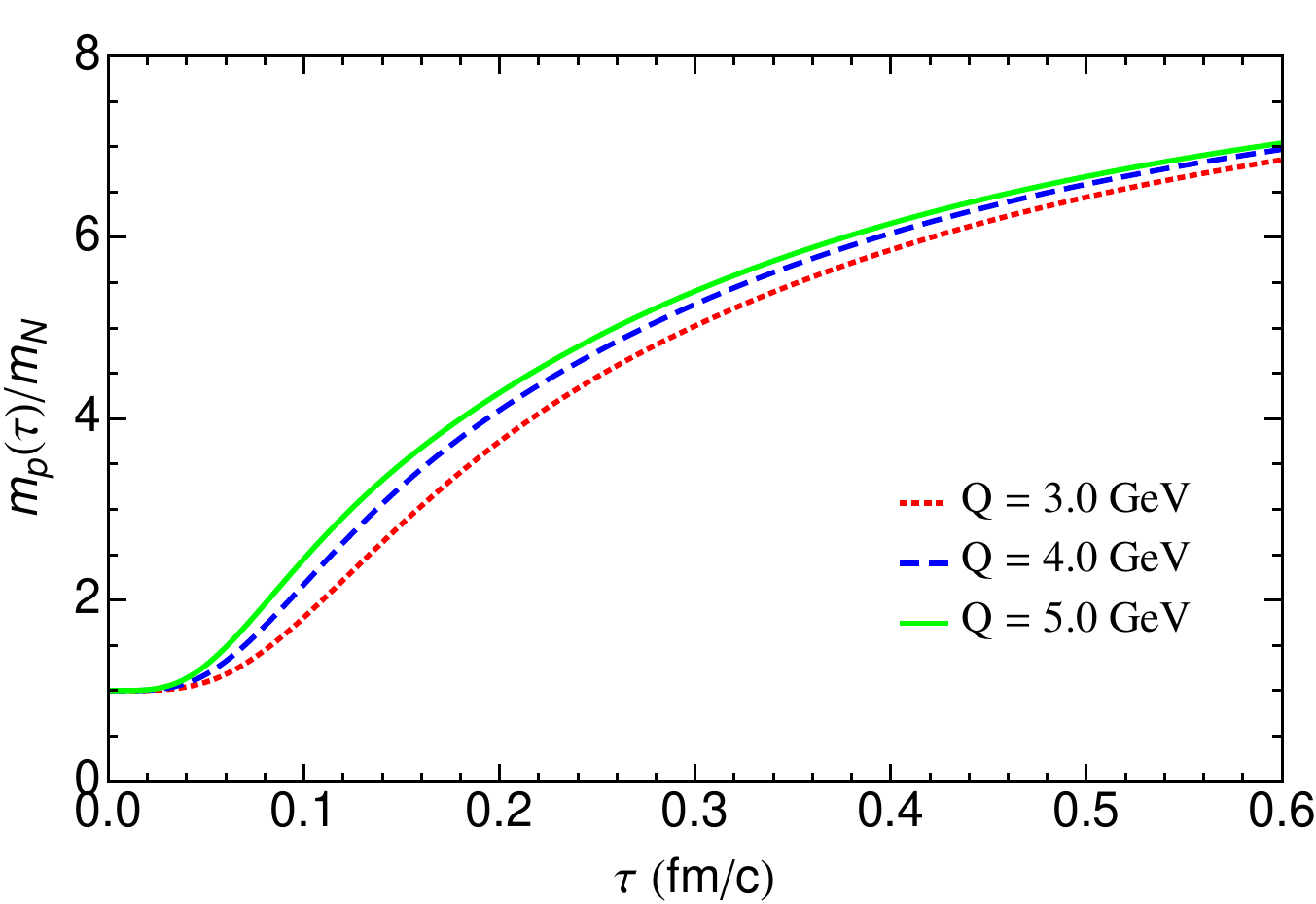}
\caption{\small Time evolution of the nuclear excitation energy scaled by the nucleon rest mass for the central core of the nucleus.}
\label{fig2_m_tau}
\end{minipage}
\end{figure}

The energy-momentum of the glasma has been calculated analytically using a small-$\tau$ power series expansion method in \cite{Chen2015, Li2016}:
\begin{equation}
T^{\mu\nu}_{\rm{glasma}}=
\begin{pmatrix}
\mathcal{A}+\mathcal{B}\cosh{2\eta} & 0 & 0 & \mathcal{B}\sinh{2\eta} \\
0 & \mathcal{A} & 0 & 0 \\
0 & 0 & \mathcal{A} & 0 \\
\mathcal{B}\sinh{2\eta} & 0 & 0 & -\mathcal{A}+\mathcal{B}\cosh{2\eta} \\
\end{pmatrix} \, .
\end{equation}
Notice that the glasma energy-momentum tensor is different from the energy-momentum tensor in the conventional string models \cite{Mishustin2002} in the sense that off diagonal terms are nonvanishing. This is due to contributions from the transverse color electric and magnetic fields represented by $\mathcal{B}$. Using the relations $v_p = dz_p/dt =\tanh y_p$, $\mathcal{E}_P =\mathcal{M}_P\cosh y_p$ and $\mathcal{P}_P = \mathcal{M}_P\sinh y_P$, the two energy-momentun conservation equations are converted to the time evolution equations for  $\mathcal{M}_P(\tau)$ and $y_P(\tau)$. To solve these equations, we need to specify the momentum space cut-off scale  Q and the initial energy density distribution $\varepsilon(r_{\perp},\tau=0)$.  The former is treated as a varying parameter ranging from $3\,\rm{GeV}$ to $ 5\,\rm{GeV}$. The latter is determined by first matching the energy density at the central region to $\varepsilon(r_{\perp}=0,\tau =0.6\,\,\rm{fm/c}) = 30\,\, \rm{GeV/fm}^3$ \cite{Song2011} that is used as initial condition for the viscous hydrodynamics. Then we assume $\varepsilon(r_{\perp})/\varepsilon(r_{\perp}=0) \sim (T_A(r_{\perp})/T_A(r_{\perp}=0))^2$ at $\tau =0$ with the nuclear thickness function $T_A(r_{\perp})=\int_{-\infty}^{+\infty} dz \,\rho_A(r_{\perp}, z)$ and $\rho_A$ the Woods-Saxon distribution for nuclear density. 

From Figure \ref{fig1_y_tau}, the rapidity for the central core of a gold nucleus decreases from the initial value $5.36$ to about $2.5$ within $0.1$ to $ 0.2 \,\,\rm{fm/c}$.  
The final net rapidity loss does not depend on the momentum space cut-off scale $Q$. We can compare our calculations using the BRAHMS result \cite{BRAHMS2004} of average rapidity loss about $2.05 + 0.4/ - 0.6$ for $0$-$10\%$ centrality class. Our calculation predicts an average rapidity loss at $\tau=0.6 \,\, \rm{fm/c}$ to be about $2.4$. We would expect the subsequent hydrodynamic evolutions and baryon diffusions only change the shape of net baryon distribution while leaving the average baryon rapdity loss unchanged. Therefore, our estimation of the baryon rapidity loss are consistent with the experimental results.   For the nuclear excitation energy, Figure \ref{fig2_m_tau} exhibits a slow but monotonic increase. This is due to the dominant roles played by the transverse components of the color electric and magnetic fields after the longitudinal fields saturate the baryon rapidity loss.

\section{High Baryon Densities}
In the McLerran-Venugopalan model, the nucleus is assumed to propagate at the speed of light both before and after the collisions. Therefore the nucleus is infinitely contracted in the center-of-mass frame but not contracted in its own local rest frame. To estimate the nuclear compression in its own rest frame after the collisions,  we resort to the simple space-time model of collisions used in \cite{Anishetty1980}. In the local rest frame of the baryonic fireball after collisions, the baryon density is 
\begin{equation}
n_B(\vec{r}_{\perp},z^{\prime}) = e^{\Delta y(\vec{r}_{\perp})} \rho_A(\vec{r}_{\perp}, e^{\Delta y(\vec{r}_{\perp})} z^{\prime}).
\end{equation}
where $z^{\prime} = z -z_P(r_{\perp})$ is the longitudinal distance in the local rest frame and $\Delta y >0$ is the rapidity change. The $\Delta y$ is a Lorentz invariant quantity and can be calculated explicitly in the center-of-mass frame as shown in Figure \ref{fig1_y_tau}. The $\rho_A(\vec{r}_{\perp},z)$ is the Woods-Saxon distribution for nuclear density. 

\begin{figure}[h]
\begin{minipage}{0.5\textwidth}
\centering
\includegraphics[width=0.9\linewidth, height=0.32\textheight]{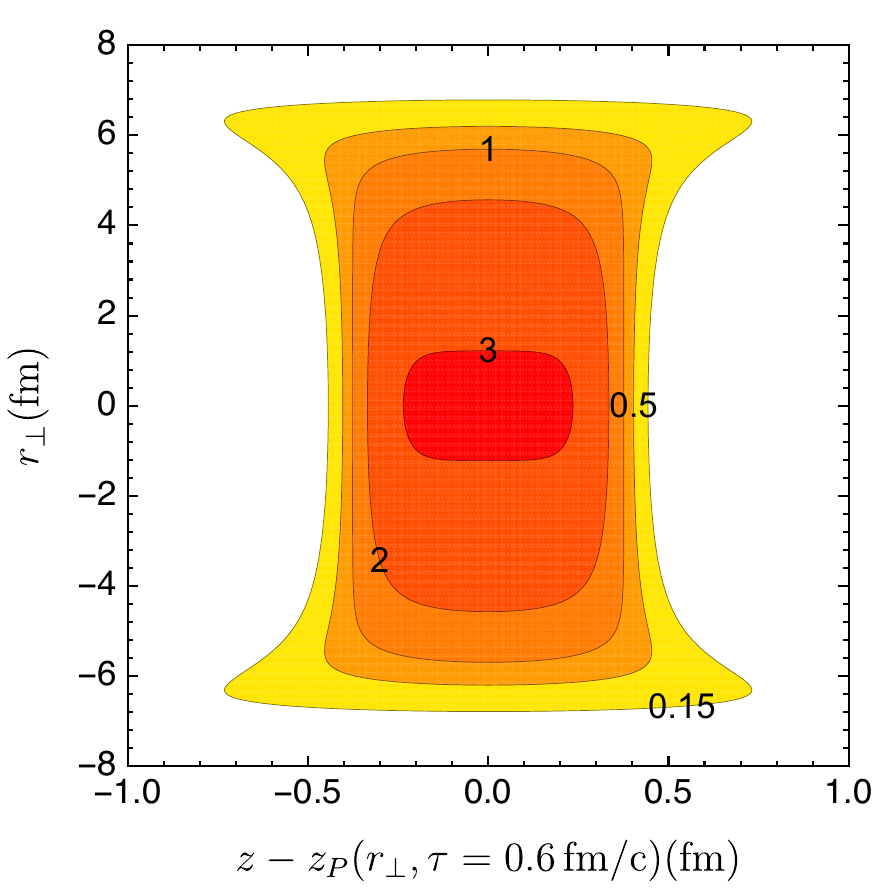}
\caption{\small Contour plot of the proper baryon density in the local rest frames of nuclear fragments. The units are baryons per $\rm{fm}^3$. The horizontal axis measures the distance in the local rest frame.}
\label{fig3_rT_z}
\end{minipage}
\hspace{0.1in}
\begin{minipage}{0.5\textwidth}
\centering
\includegraphics[width=0.9\linewidth, height=0.31\textheight]{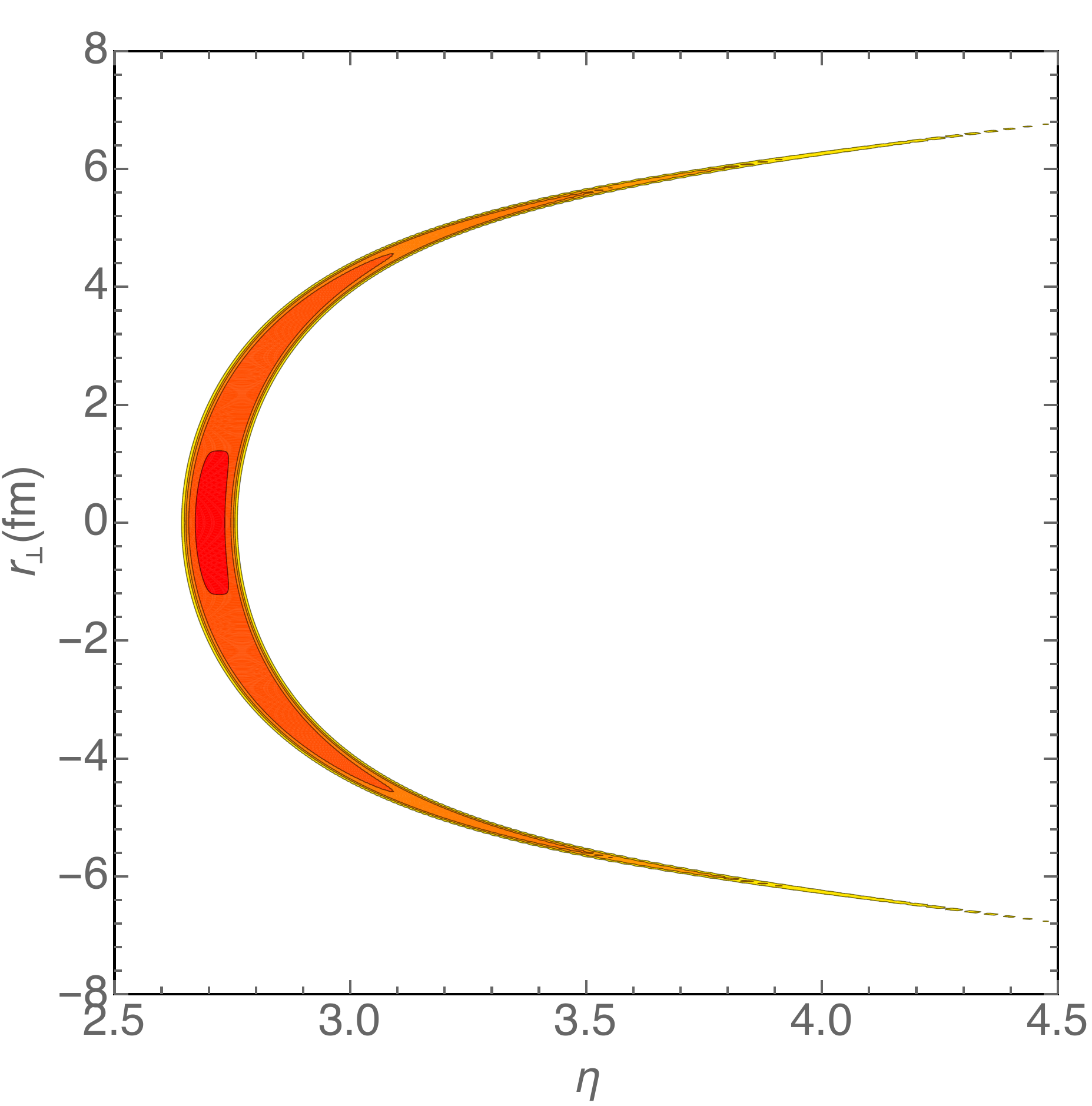}
\caption{\small Contour plot of the proper baryon density against space-time rapidity $\eta$ at $\tau = 0.6\,\, \rm{fm/c}$. The units are baryons per fm$^3$. The horizontal axis is the space-time rapidity.}
\label{fig4_rT_eta}
\end{minipage}
\end{figure}

Figure \ref{fig3_rT_z} shows the proper baryon density distribution. The contours are drawn at $n_B = 3, 2, 1, 0.5,$ and $0.15$ baryons/fm$^3$. This plot is constructed in the local rest frames of nulcear fragments which depend on different transverse coordinate $\mathbf{r}_{\perp}$. Note that the maximum baryon density can be achieved is about $3$ baryons/fm$^3$, which is about 20 times larger than the normal atomic nuclear density of $0.155$ nucleons/fm$^3$. Moreover, the longitudinal extension of the high baryon density region is about $1\,\rm{fm}$ indicating that a finite volume region of high baryon densities can be formed.  Figure \ref{fig4_rT_eta} shows the proper baryon density against the space-time rapidity $\eta$.  This is the shape of the net-baryon charge distribution at $\tau= 0.6 \,\,\rm{fm/c}$, which should be the initial distribution for the subsequent dynamical evolution. The central core region experiences the largest rapidity loss, therefore the largest baryon density and the smallest space-time rapidity.  On the other hand, the peripheral regions of the nucleus propagate with less rapidity loss, thus smaller baryon densities and larger space-time rapidities.

The local energy density distribution is readily computed with the help of the nuclear excitation energy at $\tau =0.6\,\, \rm{fm/c}$ as exampled by Figure \ref{fig2_m_tau}. However, one has to keep in mind that  the energy density obtained could be overestimated due to the monotonic increasing feature of the nuclear excitation energy.  For the central core region $r_{\perp}=0$, the net baryon density is $n_B \simeq 3.0\,\,\rm{baryons/fm}^3$ and the corresponding energy density is $\varepsilon_B \simeq20 \,\,\rm{GeV/fm}^3$. For the peripheral region, let us pick $r_{\perp} = 5.25\,\,\rm{fm}$, the baryon density is $n_B = 1.5\,\rm{baryons/fm}^3$ and the energy density is $\varepsilon_B = 5.5\,\rm{GeV/fm}^3$.  With the high baryon densities and large energy densities obtained, one is tempted to ask what the temperatures and baryon chemical potentials would be if the high baryon density matter in the fragmentation regions are thermalized. Let us approximate the high baryon density matter as a system of up, down and strange quarks and gluons. We can further assume that they are massless and noninteracting due to the large baryon density and energy density. Net strangeness is adjusted to be zero.  The equation of state thus has the form $P(T,\mu_B) = \frac{19\pi^2}{36}T^4 + \frac{1}{9}T^2\mu_B^2+\frac{1}{162\pi^2} \mu_B^4$. Typical values for the temperatures and baryon chemical potentials are $T = 299\,\,\rm{MeV}, \mu_B= 1061\,\,\rm{MeV}$ for $r_{\perp}=0$ and $T=205\,\,\rm{MeV}, \mu_B = 1007\,\,\rm{MeV}$ for $r_{\perp} = 5.25\,\rm{fm}$. The baryon chemical potentials are much larger than the temperatures indicating a very different region in the QCD phase diagram as compared to the matter created in the central rapidity region at top RHIC energy.  In addition, we obtain the entropy per baryon ratio $s/n_B = 26.2$ and  $s/n_B =18.9$, respectively.  If the following hydrodynamic expansions in the fragmentation regions are approximately adiabatic like what happened in the central rapidity region, these values would be relevant to the study of QCD phase transitions \cite{Asakawa2008}.

\section{Conclusions}
In conclusion, we used the glasma energy-momentum tensor from the McLerran-Venugopalan model to calculate both the baryon rapidity loss and the nuclear excitation energy in high energy heavy ion collisions. We obtained high baryon densities in the fragmentation regions with the largest value being 20 times larger than the normal atomic nuclear density. The initial distributions for the high baryon density and energy density in the fragmentation regions should be able to provide initial conditions for the subsequent hydrodynamic evolutions in these regions. It could also be relevant to the study of QCD phase diagram and the searching for the critical point of the phase transitions. Scanning through the large rapidity regions in high energy collisions would be equivalent to the conventional low energy beam energy scanning.

\ack{This work was supported by the U.S. Department of Energy grant DE-FG02-87ER40328}.

\end{document}